# Numerical Models for the Diffuse Ionized Gas in Galaxies

## I. Synthetic spectra of thermally excited gas with turbulent magnetic reconnection as energy source

T. L. Hoffmann, S. Lieb, A. W. A. Pauldrach, H. Lesch, P. J. N. Hultzsch, and G. T. Birk

Institute for Astronomy and Astrophysics, University of Munich, Scheinerstraße 1, D-81679 Munich, Germany

**ABSTRACT**

*Aims.* The aim of this work is to verify whether turbulent magnetic reconnection can provide the additional energy input required to explain the up to now only poorly understood ionization mechanism of the diffuse ionized gas (DIG) in galaxies and its observed emission line spectra.
*Methods.* We use a detailed non-LTE radiative transfer code that does not make use of the usual restrictive gaseous nebula approximations to compute synthetic spectra for gas at low densities. Excitation of the gas is via an additional heating term in the energy balance as well as by photoionization. Numerical values for this heating term are derived from three-dimensional resistive magneto-hydrodynamic two-fluid plasma–neutral-gas simulations to compute energy dissipation rates for the DIG under typical conditions.
*Results.* Our simulations show that magnetic reconnection can liberate enough energy to by itself fully or partially ionize the gas. However, synthetic spectra from purely thermally excited gas are incompatible with the observed spectra; a photoionization source must additionally be present to establish the correct (observed) ionization balance in the gas.

## 1. Introduction

The existence of diffuse ionized gas (DIG) in halos of disc galaxies, including the Milky Way, has been proven observationally beyond any doubt (Reynolds 1990, Rand 1996, Dettmar 1998, Rossa & Dettmar 2000, Collins & Rand 2001). From its line spectrum, characteristic values of the temperature ($10^4$ K) and number density ($10^{-1}$ cm$^{-3}$) can be deduced. Its spatial extension is significantly larger than the Strömgren spheres even of clusters of OB-stars and exhibits internal structures like bubbles, plumes, and filaments embedded in a diffuse background. Moreover, the occurrence of DIG in the halos seems to be related to the star formation rate (SFR) in the galactic discs and is associated with an extended nonthermal radio continuum and thermal X-ray halos (Dettmar 1998). The relatively high degree of polarization of these radio continuum halos (10% and higher) indicates that magnetic fields may play an important role in the origin of the DIG. Indeed, a complete solution of the ionization problem, i.e., the question of what energy sources keeps the DIG ionized, can not be given so far. On the basis of power requirement alone – a total energy input of about $10^{42}$ erg s$^{-1}$ is necessary to balance energy loss by recombination (Dettmar 1992) –, most of the ionization models, for example white dwarfs (Sokolowski & Bland-Hawthorn 1991), X-rays (McCammon et al. 1983) or cosmic rays (van Dishoeck & Black 1986), can clearly be excluded. Minter & Spangler (1997) have argued that while turbulence from supernova explosions may be energetically capable of supplying the energy that the DIG loses through radiative cooling, it is unclear how the generated turbulence could be widely distributed throughout the ISM. Likewise, photoionization by O and B stars could achieve the required power, but it is not evident how they should ionize the DIG high above the galactic plane, since the optical depth of the Lyman continuum is large. Modified photoionization models like the decaying dark matter theory (Salucci & Sciama 1991) or low-density low-excitation models (Domgörgen & Mathis 1994) are not able to explain the measured line ratios in detail without some additional heating source. By means of line ratio analysis Reynolds, Haffner, & Tufte (1999) concluded that beside O and B stars an additional heat source in the diffuse ionized gas is required. As a characteristic number they state that a density independent heating rate of some $10^{-27}$ erg cm$^{-3}$ s$^{-1}$ is required. The objective of the present work is to investigate whether magnetic reconnection can supply such an additional heating rate, and whether synthetic spectra computed for gas under these conditions can account for the observed spectra.

## 2. Synthetic spectra

Key to understanding the physical conditions in the diffuse ionized gas is the modelling of their line spectra (such as the typical spectrum shown in Fig. 1), where the models allow numerical "experiments" regarding individual microphysical processes to be performed to estimate their influence on the gas and the emitted radiation.

We will assume in this paper that the gas is homogeneous and that the ionized regions are spherically symmetric. This is of course a very crude assumption, but for a number of reasons we nevertheless think the approximation is expedient in this exploratory first step. First, at this stage we are more interested in a qualitative description of the basic ionizing mechanisms of the gas than in a detailed analysis of any particular observed ionized region, and the basic mechanisms of ionization are not seriously affected by inhomogeneities.

Second, while observations (e.g., comparing an emission measure proportional to the mean of the square of the density to a dispersion measure proportional to the mean density) do indi-cate that the gas is "clumpy", with filling factors of $f \approx 0.4\ldots1$ far away from the galactic midplane to $f \approx 0.1\ldots0.2$ in the disk (Reynolds 1991), models assuming clumping on a scale below the grid-resolution are still unsatisfactory with regard to exploring the supply of ionizing photons to the DIG. This is because this sort of "microclumping" – where the clumps are assumed to be so small that they remain optically thin – will primarily enhance the recombination rate and thus yield geometrically smaller Strömgren spheres for a given mean density (although of course the ionization balance of the metals will be analogously affected, leading to possibly different emission line strength ratios).

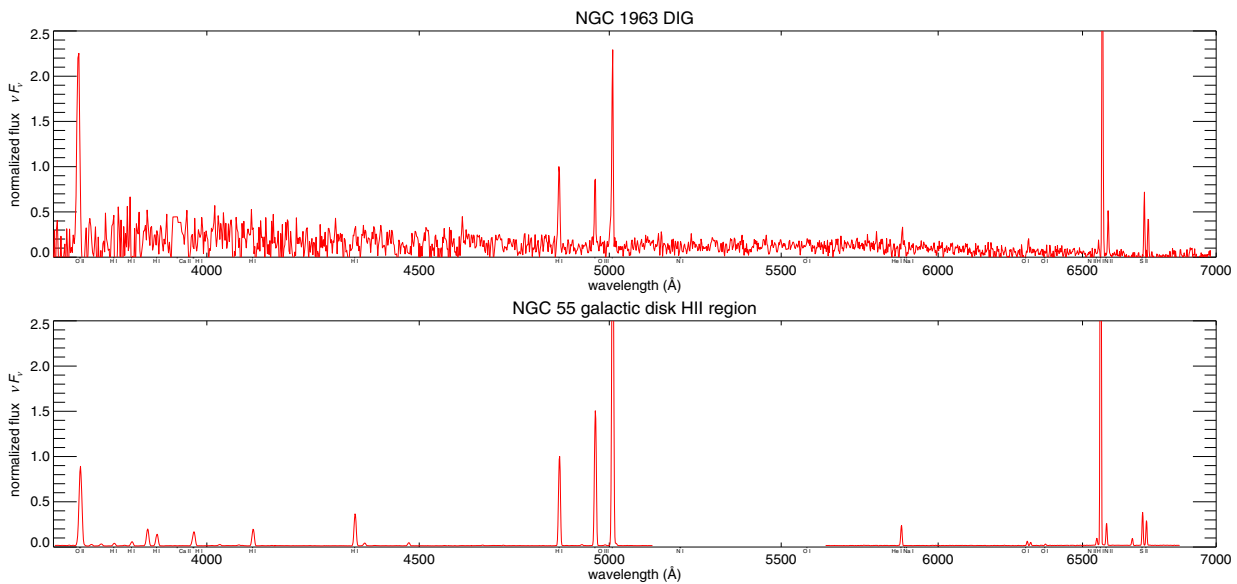

**Fig. 1.** Typical spectrum of the diffuse ionized gas (upper panel) from a region 260 pc above the galactic plane of NGC 1963 compared to that of a "classical" H II region (lower panel) from the galactic disk of NGC 55. Shown is $\nu$ times the measured flux $F_\nu$ (normalized to H$\beta$) vs. the wavelength.

Only when the geometric size of the clumps becomes significant on the scale of the H II region itself (e.g., Wood et al. 2005), and the coverage factor of the (optically thick) clumps on that scale around the source remains appreciably below 1 (Giammanco et al. 2004), will there be a significant escape of photons to regions much farther away from the ionizing source than the classical Strömgren radius corresponding to the same mean density. Indeed, simplified models for the ionization of the diffuse gas in galaxies which assume that a certain fraction of the ionizing photons escape from the observed H II regions lead to fairly good agreement with the observed distribution of the ionized gas (Zurita et al. 2002) and thus support the notion of "leaky" H II regions. Furthermore, Wood et al. (2010) conclude from hydrodynamical simulations (but without magnetic fields) that supernova-driven turbulence in the interstellar medium can render the neutral gas porous enough to allow ionizing photons from sources close to the galactic plane to escape to large galactic heights. This clear connection between H II regions and the DIG is also indicated by our current work, which shows that photoionization is essential to produce the observed distribution of the ionization stages.

However, selfconsistent photoionization models attempting to treat inhomogeneities on such scales need a fully general treatment of the radiative transfer in three dimensions. Simplified three-dimensional radiative transfer codes exist (see, e.g, Iliev et al. 2006 and references therein), but because of limits in the available computational processing power these must still make fundamental approximations with regard to the physics involved, such as the treatment of the diffuse radiation field resulting from recombination in the Lyman continuum, which is usually only considered in the "on-the-spot" approximation (cf. Iliev et al. 2006), whereas the considerable savings in computational time possible through the assumption of spherical symmetry allow the diffuse radiation field to be treated correctly. For our current work we have opted for models with spherical symmetry, but we intend to explore the use of a cartesian radiative transfer code in the future.

To calculate synthetic spectra we use a modified version of the stellar atmosphere code WM-basic, a general non-LTE radiative transfer code for computing a selfconsistent solution to ionization balance, atomic level populations, and radiative transfer, and for computing synthetic spectra from that solution. Microphysical processes considered are radiative and collisional ionization and recombination (from/to all levels), radiative and collisional excitation and de-excitation, and dielectronic recombination (where relevant). The temperature is computed from the requirement that the microphysical heating and cooling rates are balanced.

Here we briefly review the physics treated by the model code and describe the modifications we have implemented for treating gas at low densities. For specifics on the algorithms we refer the reader to Pauldrach et al. 1994, Pauldrach et al. 2001, and Pauldrach et al. 2011, where the code is described in more detail.

### 2.1. Radiative transfer

The geometry we adopt for solving the radiative transfer is illustrated in Fig. 2. We generally assume spherical symmetry for the problem, based on the concept of an idealized H II region ("nebula") ionized from a central radiation source. Radial variation of physical quantities is sampled at a sufficient number of radial grid points. Since opacity and emissivity vary strongly in the recombination region of the Strömgren sphere, the grid points are spaced more closely at that radius to adequately resolve the variation of these quantities in the radiative transfer.

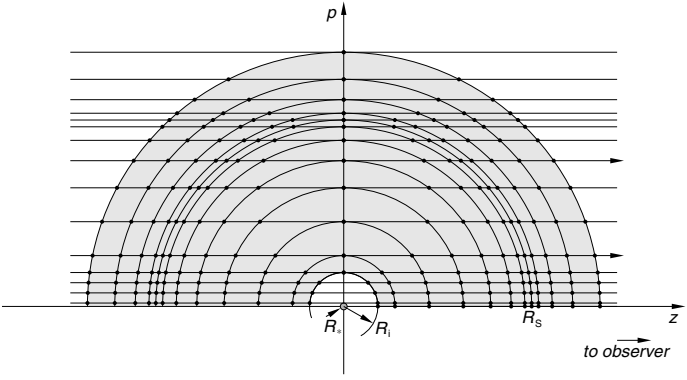

**Fig. 2.** Sketch of the $p,z$-geometry used for computing the radiative transfer and the synthetic spectra. An inner region of radius $R_i$ is considered empty of gas and therefore completely transparent. Radius grid points are set more densely at the radius $R_S$ of the hydrogen Strömgren sphere. $R_*$ is the radius of a central ionizing source.

The radiative transfer is computed along a number of "$p$-rays" with different impact parameters parallel to the central ray along the observer's line of sight. The different rays intersecting any particular radius shell describe the angular variation of the intensity at that radius, from which mean intensity $J_\nu = \frac{1}{2}\int_{-1}^{1} I_\nu\, d\mu$ and flux $H_\nu = \frac{1}{2}\int_{-1}^{1} I_\nu \mu\, d\mu$ are computed as a weighted sum over the contributions from the different angles. The diffuse radiation field from recombination and electron scattering is thus described consistently, and a-priori assumptions such as "case A" or "case B", resp. the on-the-spot approximation for the diffuse radiation, need not be made. Illumination of the nebula from the central ionizing source occurs through the central ray in the radiative transfer. However, radiation may subsequently be scattered or absorbed and reemitted, thus leading to considerable intensities also along the other $p$-rays. We allow for an inner "cavity" of (arbitrarily specifiable) radius $R_i$ that is considered empty of gas and therefore transparent.

The radiative transfer problem is treated in two steps. First, we solve the transfer equation using a Rybicky/Feautrier scheme (finite-difference method), considering all relevant continuum opacities and emissivities (bound-free, free-free, and electron scattering). Since we do not need to consider Doppler shifts in detail in this case (line transfer is treated approximately in this first step), the transfer equation is solved for only one quadrant, and a symmetric boundary condition is imposed at $z = 0$ (except for the central ray, for which an incident intensity corresponding to the spectral energy distribution of the central source is specified). The rate equations (see Sec. 2.2) are solved alternately with the radiative transfer, yielding new opacities and emissivities, and the model is iterated to convergence.

In a second step, the line transfer is also solved. Here we allow the possibility of Doppler-shifted line radiation (we assume a small expansion velocity of the nebula on the order of about 20 km/s), thus the radiative transfer has to be solved along each ray in both quadrants, i.e., through the entire nebula from the back side (moving away from the observer) to the front side (moving toward the observer). We employ an integral formulation of the transfer equation, using a spatial microgrid capable of resolving the (Doppler-shifted) line profiles. This results in the final synthetic spectrum.

The opacities and emissivities in the radiative transfer are computed as follows. The opacity associated with a bound-free transition from a level $i$ to a level $\kappa$ (in the next-higher ionization stage) is

$$\chi_{i\kappa}(\nu) = \alpha_{i\kappa}(\nu)\left(n_i - n_\kappa \left(\frac{n_i}{n_\kappa}\right)^* e^{-h\nu/kT}\right), \quad (1)$$

and the emissivity from the corresponding recombination process is

$$\eta_{i\kappa}(\nu) = n_\kappa \alpha_{i\kappa}(\nu) \left(\frac{n_i}{n_\kappa}\right)^* \frac{2h\nu^3}{c^2} e^{-h\nu/kT}, \quad (2)$$

where $n_i$ and $n_\kappa$ are the level populations, $(n_i/n_\kappa)^*$ is the Saha-Boltzmann factor, and $\alpha_{i\kappa}(\nu)$ is the cross section (see below).

The opacity and emissivity from free-free transitions (Bremsstrahlung) of a particular ionic species is

$$\chi_{\mathrm{ff}}(\nu) = n_e n^+ \alpha_{\mathrm{ff}}(\nu, T)\left(1 - e^{-h\nu/kT}\right), \quad (3)$$

$$\eta_{\mathrm{ff}}(\nu) = n_e n^+ \alpha_{\mathrm{ff}}(\nu, T) \frac{2h\nu^3}{c^2} e^{-h\nu/kT}, \quad (4)$$

where $n^+$ is the number density of the ions and $n_e$ is the free-electron density. The "cross section" $\alpha_{\mathrm{ff}}$ is computed from tabulated gaunt-factors $g_{\mathrm{ff}}$ as

$$\alpha_{\mathrm{ff}}(\nu, T) = \frac{4e^6 Z^2}{3ch}\left(\frac{2\pi}{3kTm_e^3}\right)^{1/2} \nu^{-3} g_{\mathrm{ff}} \quad (5)$$

where $Z$ is the ionic charge.

The opacity from Thomson scattering is independent of frequency and is given by

$$\chi_{\mathrm{Th}}(\nu) = n_e \sigma_{\mathrm{Th}} \quad (6)$$

with the Thomson scattering cross-section $\sigma_{\mathrm{Th}} = 6.652 \times 10^{-25}$ cm$^2$; the Thomson emissivity (assuming complete redistribution over angles) is

$$\eta_{\mathrm{Th}}(\nu) = \chi_{\mathrm{Th}} J_\nu \quad (7)$$

where $J_\nu$ is the average intensity.

Finally, the opacities and emissivities for bound-bound (line) transitions are

$$\chi_{ij}(\nu) = \frac{h\nu_{ij}}{4\pi}(n_i B_{ij} - n_j B_{ji}) \phi_{ij}(\nu - \nu_{ij}(1 + \mu v/c)) \quad (8)$$

$$\eta_{ij}(\nu) = \frac{h\nu_{ij}}{4\pi} n_j A_{ji} \phi_{ij}(\nu - \nu_{ij}(1 + \mu v/c)) \quad (9)$$

where $\nu_{ij}$ is the transition frequency and $B_{ij}$, $B_{ji}$, and $A_{ji}$ are the Einstein coefficients of the transition. For the line profile function $\phi_{ij}(\nu)$ we assume pure Doppler broadening, considering the thermal velocities of the ions and a microturbulent velocity which in this work we assume to be 10 km/s. The term $\nu_{ij}(1 + \mu v(r)/c)$ represents the projected Doppler shift of the line at any particular point $(p, z)$ in the nebula ($\mu = z/r$ and $r^2 = p^2 + z^2$).

## 2.2. Rate equations

To determine the opacities and emissivities for the radiative transfer, the occupation numbers of all contributing atomic levels have to be known. Since the densities in gaseous nebulae are so small that collisions can not establish LTE[1] against the dominating radiative processes, the occupation numbers must be explicitly computed considering all relevant atomic transitions. (In LTE, the occupation numbers simply follow a Boltzmann distribution at the local kinetic temperature.) Instead of making the usual so-called nebular approximation for thin gases (i.e., photoionization only from the ground state is considered, whereas recombination occurs to all levels, with subsequent bound-bound cascades back to the ground level), we solve the full rate equation system for all levels, thus providing a very general mechanism to compute occupation numbers applicable to all densities.

Given stationarity, the sum of all rates $n_i P_{ij}$ depopulating any particular level $i$ must be balanced by the sum of all transitions $n_j P_{ji}$ populating that level (statistical equilibrium):

$$n_i \sum_{j \neq i} P_{ij} = \sum_{j \neq i} n_j P_{ji} \tag{10}$$

where $n_i$ is the occupation number (number density) of atoms in state $i$ and $P_{ij}$ is the rate coefficient giving the probability per unit time of a transition from level $i$ to level $j$. The summations are carried out over all levels of all considered ionization stages, including excitation and de-excitation processes ($i$ and $j$ within the same ionization stage) as well as ionization and recombination processes ($i$ and $j$ in different ionization stages). The redundant system of equations is closed by replacing one of the equations by the requirement of particle conservation, $\sum_i n_i = n$, where $n = \rho Y/A$ is the number density of the element in question ($\rho$ is the local mass density, and $A$ and $Y$ are the atomic mass and abundance (mass fraction) of the element). Finally, the electron density is computed by summing over the contributions of the ionization stages of all elements considered (charge conservation).

The transition probabilities $P_{ij}$ comprise radiative ($R_{ij}$) and collisional ($C_{ij}$) transitions. For clarity, we will denote the upper level in ionization and recombination processes (bound-free transitions) by $\kappa$. This is usually the ground state of the next-higher ionization stage plus a mean over the free-electron distribution. (Where the ionization cross-sections to excited states of the next ionization stage are also relevant, these are considered as well.)

The photoionization rate coefficients are computed as a numerical integral over the radiation field intensity

$$R_{i\kappa} = \int_{\nu_{i\kappa}}^{\infty} \frac{4\pi \alpha_{i\kappa}(\nu)}{h\nu} J_\nu \, d\nu \tag{11}$$

where $\nu_{i\kappa}$ is the ionization threshold frequency and the cross sections $\alpha_{i\kappa}(\nu)$ are parametrized in this work in the "Seaton approximation"

$$\alpha_{i\kappa}(\nu) = \alpha_{i\kappa}\left(\beta_{i\kappa}\left(\frac{\nu}{\nu_{i\kappa}}\right)^{-s_{i\kappa}} + (1-\beta_{i\kappa})\left(\frac{\nu}{\nu_{i\kappa}}\right)^{-s_{i\kappa}-1}\right) \tag{12}$$

with tabulated values for $\alpha_{i\kappa}$, $\beta_{i\kappa}$, and $s_{i\kappa}$ fitted to best reproduce the cross-sections from detailed atomic calculations. Resonances in the cross-sections (autoionization, dielectronic recombination) are treated as line transitions to the ionized state $\kappa$, in a similar manner as described by Mihalas & Hummer (1973). The radiative recombination rate coefficient is computed analogously as [2]

$$R_{\kappa i} = \left(\frac{n_i}{n_\kappa}\right)^* \int_{\nu_{i\kappa}}^{\infty} \frac{4\pi \alpha_{i\kappa}(\nu)}{h\nu} \left(\frac{2h\nu^3}{c^2} + J_\nu\right) e^{-h\nu/kT} \, d\nu \tag{13}$$

where $g_i$ and $g_\kappa$ are the statistical weights of the two levels and

$$\left(\frac{n_i}{n_\kappa}\right)^* = n_e \frac{g_i}{g_\kappa} e^{h\nu_{i\kappa}/kT} \frac{1}{2}\left(\frac{h^2}{2\pi m_e kT}\right)^{3/2} \tag{14}$$

is the Saha-Boltzmann factor, the ratio of the occupation numbers that would be reached in LTE.

For collisional ionization we use an approximate formula from Seaton as given by Mihalas (1978),

$$C_{i\kappa} = n_e \frac{1.55 \times 10^{13}}{T^{1/2}} \bar{g} \alpha_{i\kappa} \frac{e^{-h\nu_{i\kappa}/kT}}{h\nu_{i\kappa}/kT} \tag{15}$$

where $\alpha_{i\kappa}$ is the photoionization cross-section at the ionization threshold and $\bar{g}$ takes the value 0.1, 0.2, or 0.3 for initial ionic charge of 0, 1, or $\geq 2$, respectively. The corresponding collisional recombination rate coefficient is

$$C_{\kappa i} = \left(\frac{n_i}{n_\kappa}\right)^* C_{i\kappa}. \tag{16}$$

However, being a three-particle process, the collisional recombination rate tends to be very small compared to the radiative recombination rate.

The rate coefficients for bound-bound (line) transitions are given by

$$R_{ij} = B_{ij} \bar{J}, \tag{17}$$

$$R_{ji} = A_{ji} + B_{ji} \bar{J}, \tag{18}$$

where we compute $\bar{J}$, the mean intensity in the line, using the Sobolev-with-continuum formalism (Hummer & Rybicki 1985; Puls & Hummer 1988), which may be thought of as a "local" escape probability method. (We adopt a small velocity gradient of 5 km/s from the inside edge of the nebula to the Strömgren radius.) For the current work we have augmented our model atoms with radiative transition data for the most important forbidden lines observed in gaseous nebulae.

The rate coefficients for collisional excitation by electrons are computed from tabulated collision strengths $\Omega_{ij}(T)$ as

$$\begin{aligned} C_{ij} &= n_e \left(\frac{2\pi}{kT}\right)^{1/2} \frac{\hbar^2}{m_e^{3/2}} \frac{\Omega_{ij}(T)}{g_i} e^{-h\nu_{ij}/kT} \\ &= n_e \frac{8.629 \times 10^{-6}}{T^{1/2}} \frac{\Omega_{ij}(T)}{g_i} e^{-h\nu_{ij}/kT}. \end{aligned} \tag{19}$$

The collisional de-excitation rate coefficient is

$$C_{ji} = \left(\frac{n_i}{n_j}\right)^* C_{ij}, \tag{20}$$

---

[1] In our simulations, deviations from LTE only regard the radiation field and the atomic level populations. As elastic collisions between particles (mostly electrons) are much more common than other processes, the kinetic thermal energies here still conform to a Boltzmann distribution (i.e., a Maxwellian distribution for the velocities).

[2] Since our current hydrogen model atom contains only 10 levels, but the recombination to these 10 levels makes up only about 90% of the total hydrogen recombination (cf. Hummer 1994), we artificially augment the recombination rate to the uppermost level to include the recombination to all higher levels not included in our model atom.

with the Boltzmann factor

$$\left(\frac{n_i}{n_j}\right)^* = \frac{g_i}{g_j} e^{h\nu_{ij}/kT}. \tag{21}$$

Where no data are available for $\Omega_{ij}$, for radiatively allowed transitions we employ van Regemorter's (1962) approximate formula (as given by Mihalas 1978) relating the collision strength to the photoexcitation cross-section,

$$C_{ij} = 5.465 \times 10^{-11} n_e T^{1/2} (14.5 f_{ij}) \bar{g} \left(\frac{E_H}{h\nu_{ij}}\right)^2 \frac{h\nu_{ij}}{kT} e^{-h\nu_{ij}/kT}, \tag{22}$$

where $E_H$ is the ionization energy of Hydrogen and $\bar{g}$ takes the value 0.7 if the main quantum numbers of levels $i$ and $j$ are equal, 0.2 otherwise. For the remaining transitions we use Eq. (19) with a representative value $\Omega_{ij} = 1$ (cf. Pauldrach 1987).

Note that since the electron velocity distribution is Maxwellian, collisions tend toward establishing LTE occupation numbers.

### 2.3. Kinetic temperature and energy balance

The temperature at each point in the gas results from the requirement that the radiative energy absorbed per unit time by any particular volume element plus the energy injected by external processes into that volume element be equal to the energy emitted per unit time by that volume element. Microphysically speaking, any imbalance in the absorption/emission of energy will result in a heating resp. cooling of the gas and in the stationary case the heating and cooling rates will completely balance.

Heating (in erg/cm$^3$/s) by radiative bound-free transitions (photoionization) is given by

$$\Gamma_{ik}^R = n_i \int_{\nu_{ik}}^{\infty} \frac{4\pi\alpha_{ik}(\nu)}{h\nu} J_\nu h(\nu - \nu_{ik}) \, d\nu \tag{23}$$

which corresponds to the radiative bound-free rate with an additional term $h(\nu - \nu_{ik})$ in the integrand, since only the difference to the ionization energy will remain as kinetic energy for the released electron. The analogous expression for cooling by radiative recombination is

$$\Lambda_{ki}^R = n_k \left(\frac{n_i}{n_k}\right)^* \int_{\nu_{ik}}^{\infty} \frac{4\pi\alpha_{ik}(\nu)}{h\nu} \left(\frac{2h\nu^3}{c^2} + J_\nu\right) e^{-h\nu/kT} h(\nu - \nu_{ik}) \, d\nu. \tag{24}$$

Radiative bound-bound transitions (lines) couple the atomic internal energy to the radiation field but do not influence the kinetic energy of the free electrons. Collisional (bound-bound – including those of radiatively "forbidden" transitions – and bound-free) processes release or consume a fixed amount of kinetic energy corresponding to the excitation resp. ionization energy of the transition. Their contribution to heating and cooling of the gas is given by

$$\begin{aligned}
\Lambda_{ik}^C &= n_i C_{ik} h\nu_{ik} & \Gamma_{ki}^C &= n_k C_{ki} h\nu_{ik} \\
\Lambda_{ij}^C &= n_i C_{ij} h\nu_{ij} & \Gamma_{ji}^C &= n_j C_{ji} h\nu_{ij}
\end{aligned} \tag{25}$$

where the $C$s are the rate coefficients. The heating and cooling by free-free transitions from a particular ionic species is computed as

$$\Gamma^{ff} = n_e n^+ \int_0^\infty 4\pi\alpha_{ff}(\nu, T) J_\nu \, d\nu \tag{26}$$

$$\Lambda^{ff} = n_e n^+ \int_0^\infty 4\pi\alpha_{ff}(\nu, T) \left(\frac{2h\nu^3}{c^2} + J_\nu\right) e^{-h\nu/kT} \, d\nu \tag{27}$$

where $n^+$ is the number density of the ions and $\alpha_{ff}(\nu, T)$ is the corresponding "cross-section". Like the bound-free heating and cooling, the evaluation of these expressions by numerical integration over frequency is straightforward.

Finally, we include in the system of heating and cooling rates to be balanced an additional heating term representing the energy released by magnetic reconnection, calculated as described in the following section.

## 3. Heating and ionization by magnetic reconnection

Given that the interstellar medium in disc galaxies is of magnetized turbulent nature, both turbulence and magnetic reconnection are generic properties of astrophysical plasma–gas systems which must be considered as important agents for energy transport. It is the aim of our contribution to show that the combination of turbulence and magnetic reconnection may provide the proposed additional heating source necessary to understand the existence of ionized gas far off standard heat and photon sources like OB stars. In a way our model is an extension of the turbulent dissipation scenario presented by Minter & Spangler (1997) and our reconnection model (Birk, Lesch, & Neukirch 1998) to magnetic reconnection in turbulent system containing plasma and neutral gas.

Minter & Spangler (1997) essentially discussed two processes, heating by the damping of MHD waves and fluid-like MHD models (Higdon 1984, 1986), whereas ion-neutral collisions are considered as the main trigger of the heating mechanism. Birk, Lesch, & Neukirch (1998) investigated in their model the heating efficiency of magnetic reconnection by conversion of free magnetic energy via Ohmic dissipation for reheating the DIG.

Magnetic reconnection can take place when magnetic field lines with antiparallel directions encounter. Such a situation is very likely to occur when the magnetic field is strongly distorted, particularly in turbulent plasmas. Turbulence in the ISM is generated by supernova activity and/or winds from OB stellar associations. Due to its low density and high temperature the interstellar plasma agitated by external sources is the perfect medium to carry turbulent magnetic fields. As long as the plasma can be regarded as an ideal conductor (zero electrical resistivity), the magnetic field is strongly coupled to the plasma motion. Any distortion of the magnetic field lines is associated with an electric field which can be described by the ideal Ohm's law

$$\boldsymbol{E} + \frac{1}{c} \boldsymbol{v} \times \boldsymbol{B} = 0. \tag{28}$$

Encounters of magnetic field lines with antiparallel directions can result in an enhanced electrical resistivity, violating the ideal form of Ohm's law, i.e., the appearance of a non-zero, localized resistivity. The field lines change their topology and reconnect again. In such a reconnection zone the magnetic energy is partially dissipated via the generation of current sheets in the resistive medium (Priest & Forbes 2000). Thus, free magnetic energy is converted into heat by Ohmic dissipation and this process yields an increasing plasma temperature. Due to thermalization processes the temperature of the neutral gas rises as well, whereby the neutrals get ionized.

A turbulent plasma–neutral-gas system will contain numerous reconnection regions, i.e., current carrying filaments driven by the interaction of plasma flows from the galactic disc with the global magnetic field in the halo system. In a former paper (Birk, Lesch, & Neukirch 1998) we have calculated the heating

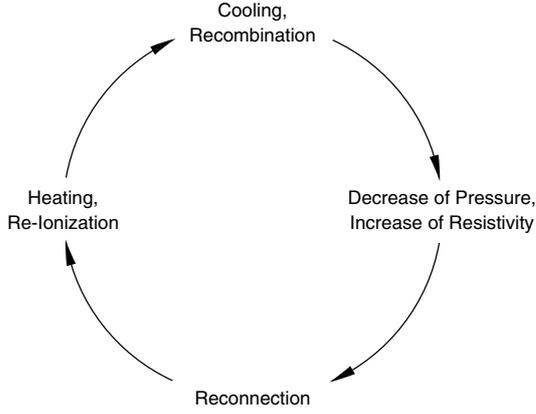

**Fig. 3.** Quasistationary ionization cycle

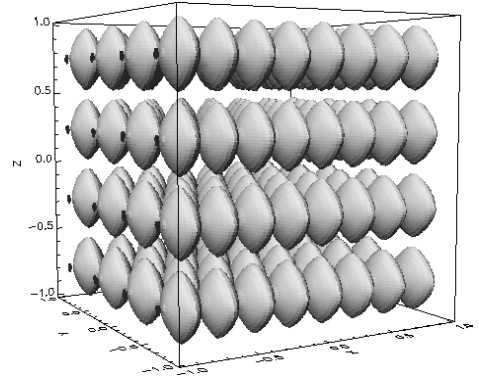

(a) Initial configuration of the magnetic field.

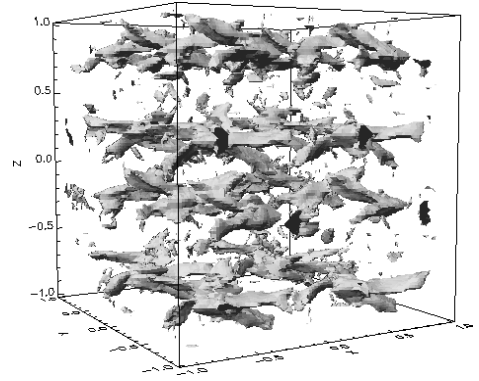

(b) Magnetic field after $t = 20$ Alfvén times.

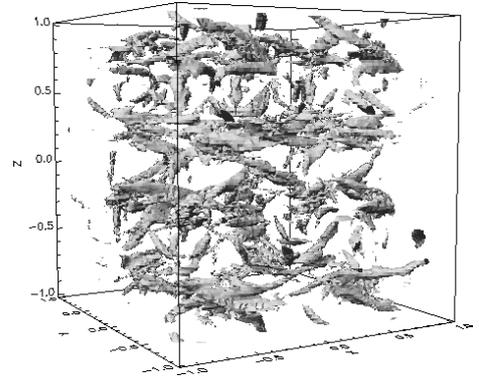

(c) Magnetic field after $t = 50$ Alfvén times.

**Fig. 4.** Configuration of the magnetic field strength $|\boldsymbol{B}|$ (shown is the isosurface corresponding to a value of 0.53 in normalized units) in a simulation with current-dependent resistivity, temperature-dependent ionization rate and constant collision frequencies. Because of the non-linearity of the system, after only 50 Alfvén times the turbulence is fully developed and the initial well ordered structure (a) is highly disarranged (c), all trace of the initial configuration being "forgotten". (After 100 Alfvén times, the turbulence has mostly decayed again, cf. Fig. 5.)

rate of one X-type reconnection event. Our main goal in this contribution is to calculate an average value of the dissipated magnetic energy into heat in some turbulent region with multiple current sheets. For these purposes we introduce three dimensional multi-fluid simulations including plasma and neutral-gas components, constant and current-dependent resistivity calculations, temperature-dependent ionization as well as recombination rates and binary collision between plasma particles and neutrals.

In this section we act on the suggestion of Birk, Lesch, & Neukirch (1998) that "heating and ionization caused by magnetic reconnection give rise to an ionized gas component". This proposal was based on a quasistationary model that is displayed in Fig. 3. Plasma far away from ordinary ionization sources like O stars recombines and cools efficiently via recombination radiation. Due to the decrease of the thermal plasma pressure, which emerge from the recombination process, the collision rate between electrons and neutrals increases. A higher rate of particle interaction leads to a lower localized electrical conductivity, hence to a higher resistivity.

Moreover, the localized drop of the thermal pressure has to balance the stresses exerted by the magnetic field which results in a contraction of the current sheet. In this situation dynamical magnetic reconnection is very likely to occur, as we know from solar flares for example (see Parker 1994 for an extended review). During the reconnection process magnetic flux is dissipated and is partially converted into heat by Ohmic dissipation. Consequently, the gas is re-heated and can be re-ionized.

In order to show that the above mentioned model indeed may account for the DIG ionization problem we have to verify the following conditions: First of all the radiative losses by bremsstrahlung and recombination radiation should be balanced by local Ohmic dissipation during the reconnection event. If the radiative losses were higher than the heating rate the DIG would recombine and cool too fast. If the opposite ratio applies the ionized halo gas would be overheated which is contradictory to the observed properties of the DIG. Thus, the energetics in our model can be described by

$$\eta j^2 = L^{\text{rad}}_{\text{brems}} + L^{\text{rad}}_{\text{recom}} \tag{29}$$

where $\eta = 1/\sigma$ ($\sigma$ is the electrical conductivity), $j$, $L^{\text{rad}}_{\text{brems}}$ and $L^{\text{rad}}_{\text{recom}}$ denote the resistivity, the current density, and the radiative loss functions due to bremsstrahlung $L^{\text{rad}}_{\text{brems}}$ and recombination radiation $L^{\text{rad}}_{\text{recom}}$, respectively.

Secondly, the reconnection process should act on time scales shorter than the recombination mechanisms otherwise the extra-planar gas would recombine totally and never be re-heated. The theory of fast reconnection (e.g. Priest & Forbes 2000) describes two different important plasma properties. One is based on a sufficiently high plasma resistivity, the other is based on sufficiently small dissipation scales. If we assume collision driven resistivity, i.e., Spitzer resistivity, the simple kind of Sweet-Parker reconnection is usually not fast enough to prevent total recombination

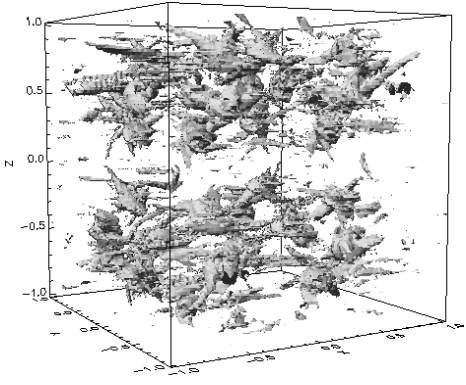

**Fig. 5.** Configuration of the magnetic field $|\boldsymbol{B}|$ after 100 Alfvén times. The turbulence has been mostly drained.

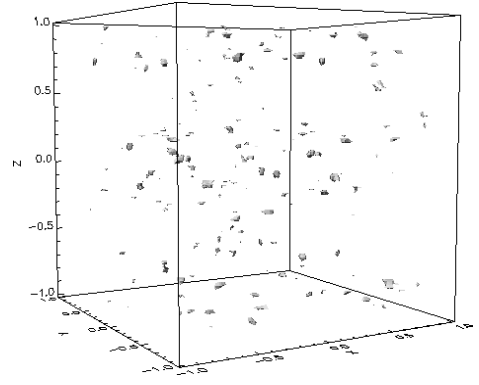

(a) Ohmic heating rate in normalized units after 20 Alfvén times.

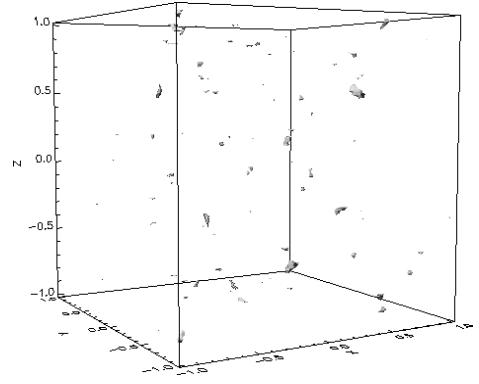

(b) Ohmic heating rate in normalized units after 50 Alfvén times.

**Fig. 6.** Ohmic heating rate: simulation with current-dependent resistivity, temperature-dependent ionization rate and constant collision frequencies.

of the DIG. The typical dissipation time in this model is (Parker 1994)

$$\tau_{\text{rec}} = \frac{\Delta}{v_A} S^{1/2} \qquad (30)$$

where $S = \Delta v_A/\eta$ is the Lundquist number ($v_A$ is the Alfvén velocity) and $\Delta$ is the width of the current sheet. Given a magnetic field strength of $B = 10^{-6}$ Gauss, a particle density of $n_p = 0.1$ cm$^{-3}$ and a plasma temperature of about $T = 1$ eV, which are characteristic values for the halo DIG, we obtain $\tau_{\text{rec}} \approx 4 \times 10^{16}$ s, if we consider the width of the current layer in the order of $10^{-9}$ pc. This time is much larger than the recombination time-scale $\tau_{\text{recom}} = 1/\alpha n_p \approx 10^{13}$ s (Dettmar 1992), where $\alpha$ is the recombination rate. However, if we include anomalous resistivities in our model, which are generated by microscopic plasma turbulence, even the relatively slow Sweet-Parker reconnection can account for the required time-scale. In addition, if the reconnection process does not operate along the entire width of the current sheet, magnetic dissipation can proceed on much shorter time-scales. A stronger localization of the dissipative scale may be caused by marginal anisotropies of recombination along the width of the current filaments. In this case magnetic reconnection is actuated by the Petschek model (Petschek 1964; Parker 1994), which operates considerably faster than the Sweet-Parker model. With the values chosen above for the magnetic field strength, the plasma density and temperature and a dissipative region $\Delta'$, which can be estimated as

$$\Delta' = \frac{\Delta (\ln S)^2}{S} = 5 \times 10^3 \text{ cm}, \qquad (31)$$

the reconnection time-scale in this model is $\tau_{\text{rec}} = 4 \times 10^4$ s. We note that the dissipation of magnetic flux in the Petschek model is fast enough to heat and thus ionize the extraplanar gas before it recombines totally.

At least, the power that is necessary to keep the DIG ionized has to provided by some external energy source. The observational correlation between the SFR in the discs of spiral galaxies and the DIG brightness in their halos indicates that disc kinetic energy is transferred in the halo gas. The general scenario suggests that on large scales the interstellar medium is characterized by energy equipartition between the dynamical constituents, i.e., cosmic rays, turbulence and magnetic fields (e.g. Ikeuchi 1988). As a result of magnetic reconnection a part of the available magnetic field is converted into heat by the above mentioned event.

If we refer to the efficiency of magnetic field dissipation as $f_{\text{diss}}$ the magnetic dissipation rate is

$$\Gamma_{\text{mag}} = \frac{B^2}{8\pi \tau_{\text{rec}}} f_{\text{diss}}. \qquad (32)$$

The energy input rate caused by disc activity is

$$\Gamma_{\text{disc}} = \frac{\dot{E}_{\text{disc}}}{V_{\text{diss}}} f_{\text{in}}, \qquad (33)$$

where $f_{\text{in}}$ denotes the fraction of transfer of disc kinetic energy to magnetic fields. $\dot{E}_{\text{disc}}$ and $V_{\text{diss}}$ are the kinetic disc luminosity and the volume of the considered region, respectively.

To be on the safe side we assume a reconnection time comparable to the recombination time of about $\tau_{\text{rec}} \approx \tau_{\text{recom}} \approx 10^{13}$ s. However, the reconnection time scale will in fact be much shorter if we include Petschek-like reconnection and/or anomalous resistivities. Furthermore we consider a sphere with a radius of 5 kpc which is considerably more than the typical scale heights of edge-on galaxies with DIG properties in their halos (Beck 1997), in particular in our galaxy. The total energy input due to disc activity into the galactic halo ranges from $10^{40}$ to $10^{42}$ erg s$^{-1}$ (Norman & Ikeuchi 1989). Since on large scales the components of the halo – gas, magnetic field, and cosmic rays – are in pressure equilibrium (Kalberla & Kerp 1999), the rate at which magnetic field is extended and free magnetic energy is continuously produced is assumed to be of the same order or a

**Table 1.** Heating rates from models with current-dependent resitivity.

| $B$ (G) | $n$ (cm$^{-3}$) | $\eta_{\text{LHD}}$ (s) | $\overline{Q}$ (erg cm$^{-3}$ s$^{-1}$) |
|---|---|---|---|
| $10^{-6}$ | 0.1 | $1.6 \cdot 10^{-8}$ | $3.3 \cdot 10^{-22}$ |
| $10^{-6}$ | 0.01 | $1.6 \cdot 10^{-7}$ | $1.9 \cdot 10^{-22}$ |
| $10^{-7}$ | 0.1 | $1.6 \cdot 10^{-9}$ | $3.2 \cdot 10^{-25}$ |
| $10^{-7}$ | 0.01 | $1.6 \cdot 10^{-8}$ | $2.1 \cdot 10^{-25}$ |
| $10^{-8}$ | 0.1 | $1.6 \cdot 10^{-10}$ | $3.0 \cdot 10^{-27}$ |
| $10^{-8}$ | 0.01 | $1.6 \cdot 10^{-9}$ | $1.9 \cdot 10^{-27}$ |

$\overline{Q}$ is the average heating rate after 50 Alfvén times, calculated by Eq. 52, for the given values of magnetic field strength and plasma density. The amount of the constant background resistivity was $10^{-13}$ for each simulation. $\eta_{\text{LHD}}$ is calculated by Eq. 45.

few magnitudes lower ($\dot{E}_{\text{disc}} f_{\text{in}} \leq 10^{41}$ erg s$^{-1}$). Hence, the input rate of free magnetic energy is

$$\Gamma_{\text{disc}} = \frac{\dot{E}_{\text{disc}}}{V_{\text{diss}}} f_{\text{in}} \leq \frac{10^{41} \text{ erg s}^{-1}}{(5 \text{ kpc})^3} \approx 10^{-25} \text{ erg cm}^{-3} \text{ s}^{-1}. \quad (34)$$

As we note in the introduction, Reynolds, Haffner, & Tufte (1999) argued for an "additional source of heat in the diffuse ionized gas". The additional heating term in their model is given by a quantity which either depends on the electron density or is just a constant factor of about $10^{-27}$ erg cm$^{-3}$ s$^{-1}$. If we compare this constant value with the value of the available energy we deduce that $f_{\text{diss}}$ should be of the order of 0.01, which indeed is a reasonable value for the case of Petschek reconnection. The magnetic dissipation rate calculated from Eq. 32 is approximately $10^{-27}$ erg cm$^{-3}$ s$^{-1}$ if we use a magnetic field strength of $5 \cdot 10^{-6}$ G and aforementioned values of $\tau_{\text{rec}}$ and $f_{\text{diss}}$. This value is consistent with the constant heating term claimed by Reynolds, Haffner, & Tufte (1999).

To summarise this section, we note that magnetic reconnection can indeed account for the required heating rate. The conversion of a tiny part of the available magnetic energy by Ohmic dissipation is an efficient heating mechanism which operates fast enough to prevent the DIG from complete recombination. The required energy is supplied by disc activity triggered by supernova activity and stellar winds. Of course, the magnetic energy would be dissipated in a large number of individual narrow current filaments, which seems to be contradictory to the observed properties of the DIG. However, as in the case of heating the solar corona by localized reconnection regions an efficient heat transport, e.g., due to shocks and outflow, on larger spatial scales can be expected (e.g., Parker 1994; Biskamp 1993). Additionally, turbulent Alfvén waves mediate between the reconnection sheets and will distribute the dissipation to spatial scales larger than the single reconnection zones (Champeaux et al. 1997). How reconnection regions "communicate" within a turbulent two fluid medium will be shown in the next section by means of three-dimensional resistive magnetohydrodynamical simulations of turbulent plasma–neutral-gas systems.

### 3.1. Numerical simulations overview

Here we introduce three-dimensional dynamical simulations of a plasma–neutral-gas system, which show that magnetic reconnection in turbulent plasma–neutral-gas systems may contribute to the ionization of the extraplanar diffuse ionized halo gas.

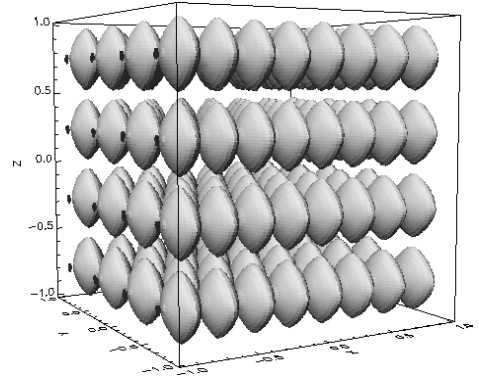

(a) Initial configuration of the magnetic field in normalized units.

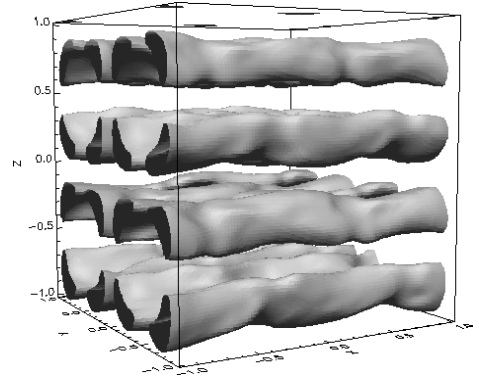

(b) Magnetic field in normalized units after $t = 20$ Alfvén times.

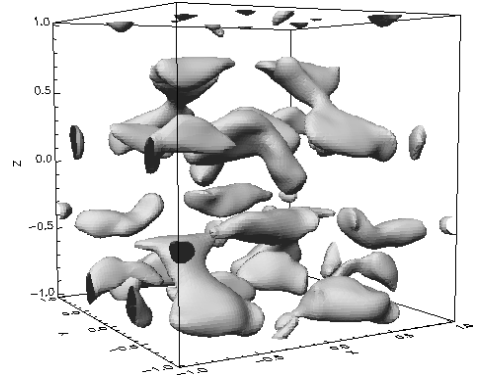

(c) Magnetic field in normalized units after $t = 50$ Alfvén times.

**Fig. 7.** Configuration of the magnetic field $|\boldsymbol{B}|$: simulation with constant resistivity and collision frequencies and temperature-dependent ionization rate.

Our general physical picture relies on the idea that magnetic reconnection is always connected to deviations from idealness, i.e., the electrical conductivity has to be reduced by some plasma instability (Schindler et al. 1991). In other words, the dependency of the reconnection process on the anomalous collision frequency is not really important. We use a model where plasma waves (in our case lower hybrid waves) represent the mechanism to reduce the conductivity. We have chosen these waves because they have the lowest threshold, namely $v_{\text{d}} > v_{\text{T}_{\text{i}}}$ (where $v_{\text{d}}$ is the electron drift velocity and $v_{\text{T}_{\text{i}}}$ is the thermal speed of the ions, see Eqs. 44 and 45 below) and they are well understood as source terms for anomalous resistivity in the context of collisionless reconnection in current sheets.

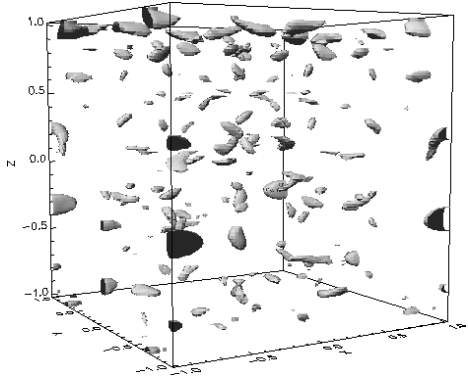

(a) Ohmic heating rate in normalized units after 20 Alfvén times.

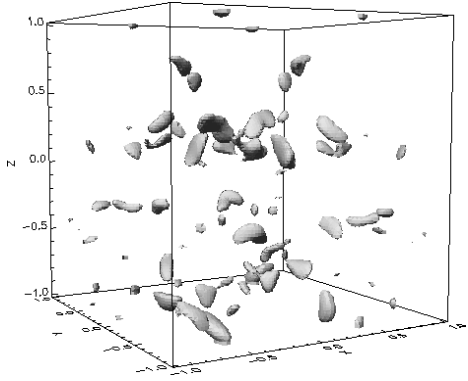

(b) Ohmic heating rate in normalized units after 50 Alfvén times.

**Fig. 8.** Ohmic heating rate: simulation with constant resistivity and collision frequencies and temperature-dependent ionization rate.

Our numerical model is meant to be an example how magnetic reconnection can be driven by the onset of turbulence. In our case we have chosen Orszag-Tang turbulence (Orszag & Tang 1979) because of its convenient mathematical properties. In general, reconnection on small scales does not depend on the large scale dynamics which unavoidably must result in strong localized magnetic fields, i.e., strong current sheets evolve in which at some stage the current density $j$ exceeds a critical value $j_{\text{crit}}$. The required numeric resolution must be high, of course, in order to resolve reconnection events. Indeed, it is one of the central challenges for nonideal MHD simulations to connect large-scale motions with small-scale events like magnetic dissipation in current sheets. Our approach is to connect them by the condition that $j > j_{\text{crit}}$, as described below.

The spatial extension of our simulation volume ($10^9$ cm in each direction, with a resolution of $101^3$ grid points) has been chosen to be of the same dimension as the inner scale of the turbulence in the DIG proposed by Minter & Spangler (1997), i.e., $10^8$ cm, in accordance with the measurements of Armstrong et al. (1995) who show that the interstellar medium is turbulent over a scale from $10^{15}$ cm down to $10^8$ cm and likely even smaller (over more than 6 orders of magnitude!) with a power-law density distribution consistent with a Kolmogorov-like power law over 10 decades. Thus, even though we may not know in detail the exact mechanism that causes the turbulence, we know that it exists and that it maintains the turbulence. In this context it is also interesting to note that this relative small volume indicates that a huge number of small current sheets is necessary in total to provide at least partly the required heating rate in the area of the ISM where diffuse ionized gas has been observed.

The balance equations of the multi-fluid system that are integrated via an explicit difference scheme (for details of the numerical code see Birk & Otto 1996) are:

$$\frac{\partial \rho}{\partial t} = -\nabla \cdot (\boldsymbol{v}\rho) + \iota\rho\rho_n - \alpha\rho^2, \tag{35}$$

$$\frac{\partial \rho_n}{\partial t} = -\nabla \cdot (\boldsymbol{v}_n\rho_n) - \iota\rho\rho_n + \alpha\rho^2, \tag{36}$$

$$\begin{aligned}\frac{\partial}{\partial t}(\rho\boldsymbol{v}) = &- \nabla \cdot (\rho\boldsymbol{v}\boldsymbol{v}) - \nabla p + (\nabla \times \boldsymbol{B}) \times \boldsymbol{B} \\ &- \rho\nu_{pn}(\boldsymbol{v} - \boldsymbol{v}_n) + \iota\rho\rho_n\boldsymbol{v}_n - \alpha\rho^2\boldsymbol{v},\end{aligned} \tag{37}$$

$$\begin{aligned}\frac{\partial}{\partial t}(\rho_n\boldsymbol{v}_n) = &- \nabla \cdot (\rho_n\boldsymbol{v}_n\boldsymbol{v}_n) - \nabla p_n \\ &- \rho_n\nu_{np}(\boldsymbol{v}_n - \boldsymbol{v}) - \iota\rho\rho_n\boldsymbol{v}_n + \alpha\rho^2\boldsymbol{v},\end{aligned} \tag{38}$$

$$\frac{\partial \boldsymbol{B}}{\partial t} = \nabla \times (\boldsymbol{v} \times \boldsymbol{B}) - \nabla \times (\eta\nabla \times \boldsymbol{B}), \tag{39}$$

$$\begin{aligned}\frac{\partial p}{\partial t} = &- \boldsymbol{v} \cdot \nabla p - \gamma p \nabla \cdot \boldsymbol{v} \\ &+ (\gamma - 1)\bigg[\eta(\nabla \times \boldsymbol{B})^2 - \nu_{pn}\bigg(p - \frac{\rho}{\rho_n}p_n\bigg) + \iota\rho p_n \\ &- \alpha\rho p + (\rho\nu_{pn} + \frac{1}{2}(\iota\rho\rho_n + \alpha\rho^2))(\boldsymbol{v} - \boldsymbol{v}_n)^2\bigg],\end{aligned} \tag{40}$$

$$\begin{aligned}\frac{\partial p_n}{\partial t} = &- \boldsymbol{v}_n \cdot \nabla p_n - \gamma_n p_n \nabla \cdot \boldsymbol{v}_n \\ &- (\gamma_n - 1)\bigg[\nu_{np}\bigg(p_n - \frac{\rho_n}{\rho}p\bigg) + \iota\rho p_n - \alpha\rho p \\ &+ (\rho_n\nu_{np} + \frac{1}{2}(\iota\rho\rho_n + \alpha\rho^2))(\boldsymbol{v}_n - \boldsymbol{v})^2\bigg].\end{aligned} \tag{41}$$

Here $\rho$, $\boldsymbol{v}$, $p$, $\nu_{pn}$, $\nu_{np}$, $\gamma$, $\gamma_n$, $\iota$ and $\alpha$ denote the mass density, the bulk velocity, the thermal pressure, the elastic plasma–neutral-gas collision frequencies ($\nu_{pn}\rho = \nu_{np}\rho_n$), the ratios of the specific heats and the ionization and recombination rates. The values of $\nu_{pn}$, $\nu_{np}$, $\iota$, $\alpha$ and $\eta$ are either kept constant or are calculated consistently by the corresponding values of the system. In our simulations $\nu_{pn}$ and $\nu_{np}$ are kept consistently constant since for our purpose the influence of a temporal variability on the results can be neglected. Simulations with momentum-dependent collision frequencies have shown no significant deviation. The exclusive role of this term consists in the thermalization of the neutral gas. The ionization and recombination rates are given by (Goldston & Rutherford 1995)

$$\iota = \frac{2 \cdot 10^{-13}}{6.0 + T_e/13.6} \sqrt{\frac{T_e}{13.6}} \exp\left(-\frac{13.6}{T_e}\right) \text{ m}^3 \text{ s}^{-1}, \tag{42}$$

$$\alpha = 0.7 \cdot 10^{-19} \sqrt{\frac{T_e}{13.6}} \text{ m}^3 \text{ s}^{-1}, \tag{43}$$

where $T_e$ denotes the electron temperature measured in eV. Of course collisions with ions can also account for ionization of the neutrals, but the velocities of ions are assumed to be much lower

**Table 2.** Heating rates from models with constant resistivity.

| $B$ (G) | $n$ (cm$^{-3}$) | $\eta_{\text{Spitzer}}$ (s) | $\overline{Q}$ (erg cm$^{-3}$ s$^{-1}$) |
|---|---|---|---|
| $10^{-6}$ | 0.1 | $3 \cdot 10^{-10}$ | $1.6 \cdot 10^{-20}$ |
| $10^{-6}$ | 1.0 | $3 \cdot 10^{-10}$ | $6.4 \cdot 10^{-23}$ |
| $10^{-7}$ | 0.1 | $3 \cdot 10^{-11}$ | $1.7 \cdot 10^{-23}$ |
| $10^{-7}$ | 1.0 | $3 \cdot 10^{-11}$ | $6.6 \cdot 10^{-26}$ |
| $10^{-8}$ | 0.1 | $3 \cdot 10^{-12}$ | $1.3 \cdot 10^{-26}$ |
| $10^{-8}$ | 1.0 | $3 \cdot 10^{-12}$ | $6.7 \cdot 10^{-29}$ |

$\overline{Q}$ is the average heating rate after 50 Alfvén times, from simulations with temperature-dependent ionization rate and constant resistivity. The Spitzer resistivities were chosen 1 to 3 magnitudes higher (cf. Sec. 3.3) than the intrinsic value of $10^{-13}$ s which corresponds to an electron temperature of 1 eV.

than the velocities of the electrons and hence ionization due to ions is neglected.

We generate turbulence by a superposition of Alfvén waves, i.e., an initially sine-like disturbance in each direction of both the magnetic field and the plasma fluid velocity. Since the ISM is a magnetized medium, turbulence in the ISM can be considered as MHD turbulence stimulated by MHD waves (Minter & Spangler 1997). As can be seen in Fig. 4 the turbulence is indeed fully developed after about 50 Alfvén times.

In the first simulation we have dealt with a current-dependent resistivity model which is based on the theory of anomalous resistivity. The basic result of this theory is that for the initiation of the resistivity the drift speed ($v_d$) between ions and electrons has to exceed the ion thermal speed ($v_{T_i}$) (Priest & Forbes 2000). This leads to a threshold value for the critical current of

$$j_{\text{crit}} = n_e e v_{T_i} \tag{44}$$

where $n_e$ and $e$ denote the electron density and electrical charge, respectively. As long as the current density does not exceed $j_{\text{crit}}$ the resistivity is dominated by Coulomb collisions. To account for Coulomb collision driven resistivity we have used a small and constant background resistivity in our simulations comparable to the value of Spitzer resistivity.

If the current exceeds $j_{\text{crit}}$, the resistivity will rise abruptly, provoked by micro-instabilities. The lower-hybrid drift instability (LHDI), for example, is driven by cross-field drifts and the diamagnetic current in the presence of density and magnetic field gradients. Local appearance of this situation is likely to occur in turbulent plasma systems. Even though the fastest growing modes of the LHDI is localized on the edge of the current sheet, resonant electron scattering in the nonlinear regime triggers the LHDI in the central region of the sheet (Daughton, Lapenta, & Ricci 2004). Furthermore, Scholer et al. (2003) have presented fully consistent three-dimensional particle-in-cell simulations with the result that the LHDI have led to the onset of collisionless fast reconnection in current sheets. The lower-hybrid drift instability is assumed to be responsible for the current-dependence of the resistivity, which is given by

$$\eta_{\text{LHD}} = \frac{4\pi \sqrt{\omega_{c_i} \omega_{c_e}}}{\omega_{p_e}^2} \tag{45}$$

where $\omega_{c_i}$ and $\omega_{c_e}$ are the ion and electron collision frequencies and $\omega_{p_e}$ is the plasma frequency.

### 3.2. Simulations with current-dependent resistivity

As a result of the MHD turbulence and the resistivity model, we obtain a spatially very localized ohmic heating rate (Fig. 6). After about 50 Alfvén times the number of regions in which magnetic flux is dissipated has been drastically reduced compared to after 20 Alfvén times. This originates from the fact that in the numerical scheme we excite the turbulence only once. Instead of an explicit source term we have used periodic boundary conditions in our simulations, i.e., there is no net flux of any field quantities through the boundaries out of the simulation box. Thus, during the simulation the turbulence energy decreases by dissipation processes and with it the current density and the ohmic heating rate. In fact, the turbulence has been nearly drained after a calculation time of 100 Alfvén times (Fig. 5).

To calculate the total average heating rate we argue as follows: We start our simulations by perturbing a homogeneous static equilibrium by non-linear interacting Alfvén waves to generate a three-dimensional generalization of the Orszag-Tang kind of turbulence (Orszag & Tang 1979), i.e.,

$$B_x = -B_0 \sin\left(2\pi \frac{y}{y_{\max}}\right) \sin\left(2\pi \frac{z}{z_{\max}}\right) \tag{46}$$

$$B_y = B_0 \sin\left(4\pi \frac{x}{x_{\max}}\right) \sin\left(2\pi \frac{z}{z_{\max}}\right) \tag{47}$$

$$B_z = B_0 \sin\left(4\pi \frac{x}{x_{\max}}\right) \sin\left(2\pi \frac{y}{y_{\max}}\right) \tag{48}$$

$$v_x = -v_0 \sin\left(2\pi \frac{y}{y_{\max}}\right) \sin\left(2\pi \frac{z}{z_{\max}}\right) \tag{49}$$

$$v_y = v_0 \sin\left(2\pi \frac{x}{x_{\max}}\right) \sin\left(2\pi \frac{z}{z_{\max}}\right) \tag{50}$$

$$v_z = v_0 \sin\left(2\pi \frac{x}{x_{\max}}\right) \sin\left(2\pi \frac{y}{y_{\max}}\right), \tag{51}$$

where $x_{\max}$, $y_{\max}$, and $z_{\max}$ are the upper limits of the numerical box, respectively. The amplitudes are chosen as $B_0 = 0.5$ and $v_0 = 0.05$ in normalized units. This excites MHD turbulence in the plasma. After the turbulence is fully developed a large number of individual current filaments has been generated where magnetic field energy is dissipated. In the case of simulations with a current-dependent resistivity the dissipation via ohmic heating sets in when the current density exceeds the critical value where the resistivity is driven by the lower hybrid drift instability. Then the total average heating rate $\overline{Q}$ derives from

$$\overline{Q} = \frac{1}{V} \int f(\eta) \, j^2 \, dV \tag{52}$$

where $f(\eta) = \eta_{\text{LHD}}$ if $j \geq j_{\text{crit}}$ and $f(\eta) = \eta_{\text{Spitzer}}$ if $j < j_{\text{crit}}$. For simplification we have used a constant value of $\eta_{\text{Spitzer}} = 3 \cdot 10^{-13}$ s for the Spitzer resistivity, which corresponds to an electron temperature of $T_e = 1$ eV. $V$ denotes the volume of the numerical box. Since our current 3d simulation does not yet model a driven turbulence, we take as a representative value of the time-averaged heating rate[3] the value after a simulation

---

[3] Konz et al. (2000) show that when mechanical energy is continuously input into such a system (in their (2d) simulation, through constant shearing) a steady-state situation develops in which energy input and dissipation (through magnetic reconnection) balance each other (their Fig. 11).

**Table 3.** Parameters of the radiative transfer models.

| Model | $n_H$ (cm$^{-3}$) | $R_*$ ($R_\odot$) | $T_*$ (kK) | $R_i$ (pc) | $R_S$ (pc) | $\overline{Q}$ (erg cm$^{-3}$ s$^{-1}$) | $n_e/n_H$ | $T_e$ (K) | $E_{ion}$ (erg s$^{-1}$) | $E_{heat}$ (erg s$^{-1}$) |
|---|---|---|---|---|---|---|---|---|---|---|
| photoionized nebulae | | | | | | | | | | |
| A4 | 10 | 19 | 40 | 1 | 21 | 0 | 1.1 | ~ 8000 | $1.3 \times 10^{39}$ | 0 |
| B4 | 1 | 19 | 40 | 1 | 99 | 0 | 1.1 | ~ 8000 | $1.3 \times 10^{39}$ | 0 |
| C4 | 0.1 | 19 | 40 | 1 | 465 | 0 | 1.1 | ~ 8000 | $1.3 \times 10^{39}$ | 0 |
| B3 | 1 | 19 | 30 | 1 | 56 | 0 | 1.1 | ~ 7400 | $2.2 \times 10^{38}$ | 0 |
| thermally excited nebulae | | | | | | | | | | |
| B4a | 1 | $19 \cdot 10^{-3}$ | 40 | 1 | 99 | $6.4 \cdot 10^{-23}$ | 0.52 | 15000 | $1.3 \times 10^{33}$ | $7.6 \times 10^{39}$ |
| B4b | 1 | $19 \cdot 10^{-3}$ | 40 | 1 | 99 | $6.6 \cdot 10^{-26}$ | 0.0045 | 10000 | $1.3 \times 10^{33}$ | $7.9 \times 10^{36}$ |
| B4c | 1 | $19 \cdot 10^{-3}$ | 40 | 1 | 99 | $6.7 \cdot 10^{-29}$ | 0.00035 | 3700 | $1.3 \times 10^{33}$ | $8.0 \times 10^{33}$ |
| C4a | 0.1 | $19 \cdot 10^{-3}$ | 40 | 1 | 465 | $1.6 \cdot 10^{-20}$ | — | — | $1.3 \times 10^{33}$ | $2.0 \times 10^{44}$ |
| C4b | 0.1 | $19 \cdot 10^{-3}$ | 40 | 1 | 465 | $1.7 \cdot 10^{-23}$ | — | — | $1.3 \times 10^{33}$ | $2.1 \times 10^{41}$ |
| C4c | 0.1 | $19 \cdot 10^{-3}$ | 40 | 1 | 465 | $1.3 \cdot 10^{-26}$ | 0.039 | 12000 | $1.3 \times 10^{33}$ | $1.6 \times 10^{38}$ |
| photoionized *and* thermally excited nebulae | | | | | | | | | | |
| B4a1 | 1 | 19 | 40 | 1 | 155 | $6.4 \cdot 10^{-23}$ | 1.1 | ~ 25000 | $1.3 \times 10^{39}$ | $2.9 \times 10^{40}$ |
| B3a1 | 1 | 19 | 30 | 1 | 85 | $6.4 \cdot 10^{-23}$ | 1.1 | ~ 25000 | $2.2 \times 10^{38}$ | $4.8 \times 10^{39}$ |

$R_*$ and $T_*$ refer to the radius and the effective temperature of the central ionizing source, $n_H$, $R_i$, and $R_S$ to the total hydrogen density of the nebula, the radius of the inner nebular "cavity", and the Strömgren radius of the ionized bubble. $\overline{Q}$ is the additional heating rate, $n_e/n_H$ the resulting relative electron density of the gas, and $T_e$ the resulting nebular equilibrium temperature. $E_{ion}$ is the total ionizing energy emitted by the central source per unit time, and $E_{heat}$ the total energy per unit time deposited via heating within the ionized region.

time of 50 Alfvén times, at which time the turbulence is fully developed (see Fig. 4) and before it has decayed again (which is the case at about 100 Alfvén times, see Fig. 5). The heating rates calculated after 50 Alfvén times for different magnetic field strengths and plasma densities are given in Table 1.

The main results are: The total average heating rate is almost independent of the electron density, i.e., $\overline{Q}$ is primarily determined by the dissipated magnetic field energy. Measurements of the mean magnetic field in the Reynolds layer by Minter & Spangler (1996) give a characteristic field strength of ~ 3 $\mu$G. This means that only a fraction of less than one percent of the available magnetic field has to be dissipated by magnetic reconnection in order to obtain the supplemental heating rate of a few times $10^{-27}$ erg cm$^{-3}$ s$^{-1}$ claimed by Reynolds, Haffner, & Tufte (1999).

On the other hand, if most of the magnetic flux available was involved in the reconnection processes, the heating rate would even exceed the maximum of the interstellar cooling function (Dalgarno & McCray 1972), which would allow heating the extraplanar gas up to the X-ray range. Whether turbulent magnetic reconnection on its own can explain the X-ray properties of the ionized halo gas in spiral galaxies is questionable, but in the direct vicinity of the current sheets this situation is very likely to occur.

### 3.3. Simulations with constant resistivity

Although the scenario with constant resistivity is less realistic than the current-dependent case, we want to present these numerical results to show that our model could even work without anomalous resistivity. The difficulty in this case is handling the quite low values of the Spitzer resistivity in our numerical scheme. In order to avoid numerical problems occurring with very low resistivities, we have used resistivities 1 to 3 magnitudes higher than the intrinsic value. This constraint should not falsify the values of the ohmic heating rate significantly, since a higher resistivity gives rise to a lower current density (cf. Eq. 29).

Compared to the results of the simulations with current-dependent resistivity the turbulent magnetic field after 50 Alfvén times (Fig. 7) is more smooth and the regions of the ohmic heating rate with significant amplitudes (Fig. 8) are more numerous and much broader. As opposed to the results in Sec. 3.2 the average heating rate depends on plasma density and magnetic field strength as it is given in Table 2. The values of $\overline{Q}$ are even higher than in the case of simulations with anomalous resistivity. This is based on the fact that the resistivity was chosen somewhat higher than the background resistivity in Sec. 3.2 and that the development of the turbulence proceeds considerably slower and therefore more magnetic field energy is available than in the runs with a current-dependent resistivity (compare Fig. 4 and Fig. 7). Indeed, the heating rates in Table 2 are approximately two magnitudes lower after a simulation time of 100 Alfvén times. Overall, the results match the findings of the previous subsection.

## 4. Results

Using the heating rates from the magnetic reconnection simulations we have carried out radiative transfer simulations to investigate how the additional heating term from magnetic reconnection will influence the spectra of the gas. We have used the values from Table 2 but we note that the specific numbers are not crucial as the heating rate can, depending on the physical conditions in the gas (density, magnetic field strength), span several orders of magnitude. Table 3 lists the parameters for a selection of models we have computed. The models cover the range of expected magnetic field strengths (resp. the average heating rates derived from these) and gas densities. All models represent a spherical ionized/excited bubble in a constant-density (save for the inner cavity) environment. We use solar values (Grevesse & Sauval 1998) for the abundances.

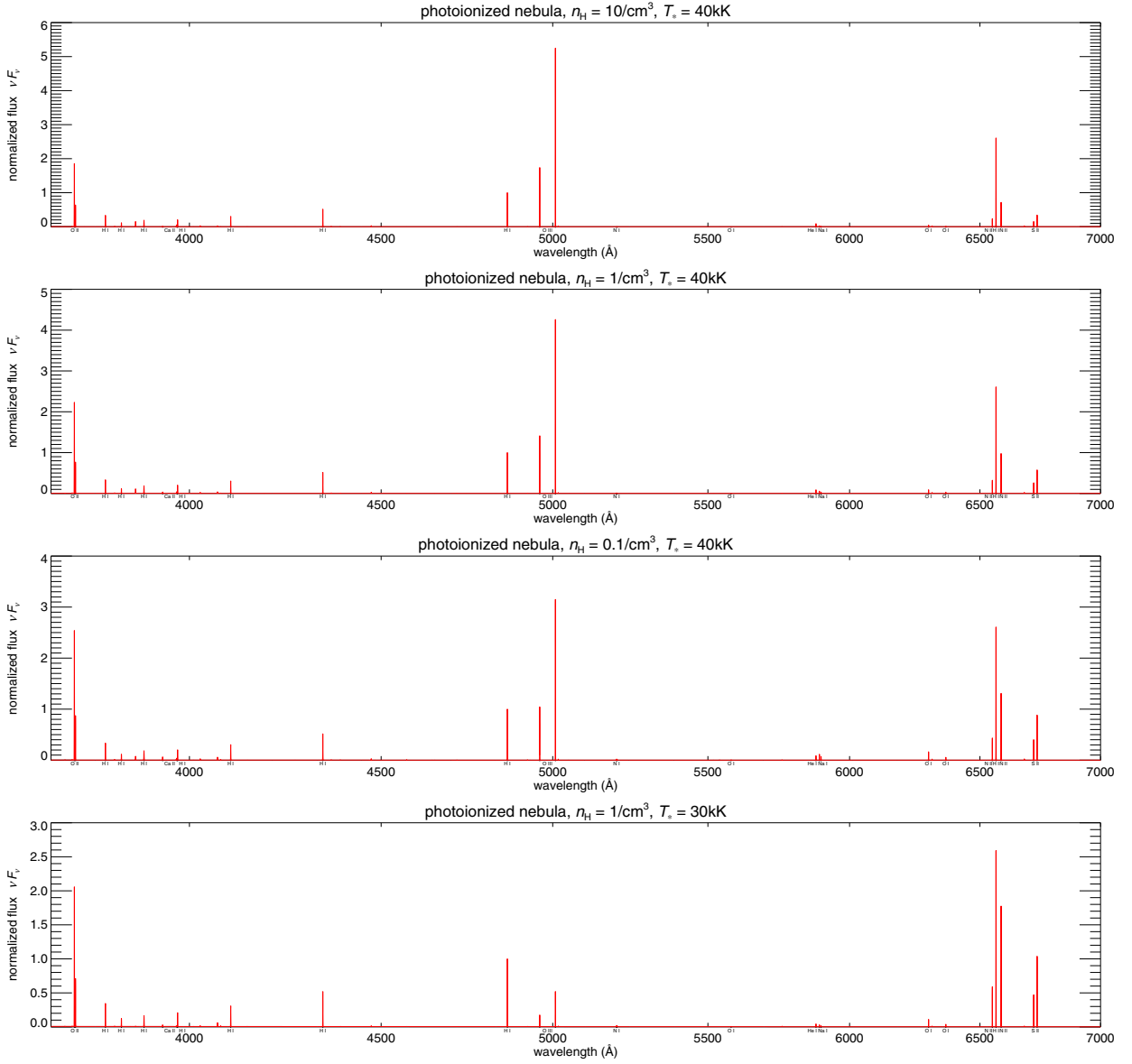

**Fig. 9.** Spectra of models heated through photoionization only, normalized to H$\beta \equiv 1$. (Models A4, B4, C4, B3 (from top to bottom); the parameters of the models are listed in Table 3).

### 4.1. Models heated by photoionization

Besides the models including the additional heating term we have also computed a number of purely photoionized models for comparison purposes. Their synthetic spectra are plotted in Fig. 9. They exhibit the well-known characteristics of H II region spectra, although our models predict a somewhat strong [O III] emission. ([O III] $\lambda$ 4959,5007 assumes the role of major coolant in these models since our current atomic model does not yet consider fine-structure transitions and the cooling via collisional excitation and radiative emission in fine-structure lines is therefore absent.) Furthermore, as our aim in computing these models has at this point not been as a diagnostic means to analyze any particular observed spectrum, but rather to compare different excitation mechanisms, we have elected to simply use a black body as central ionizing source instead of, more realistically, the emergent flux from a model stellar atmosphere. Nevertheless, these models do illustrate the range of line-strength ratios that can be expected from a variation of gas density and "hardness" of the stellar ionizing radiation. (Note, however, that all these models are radiation-bounded. We have not considered the possibility of an "open" geometry where part of the ionizing radiation may freely escape.)

### 4.2. Thermally excited models

To make the thermally exited models comparable to the photoionized models, in particular with regard to the energy budget, we have attempted to keep the geometry as similar as possible. Thus, we restrict heating to a region of the size of the Strömgren sphere of the corresponding photoionized model, and reduce the luminosity of the central source by a factor of $10^6$, thereby effectively disabling photoionization. (We have confirmed this by verifying that the physical properties at the innermost grid point of the converged model differ only marginally from the properties of the grid points farther away from the central source.)

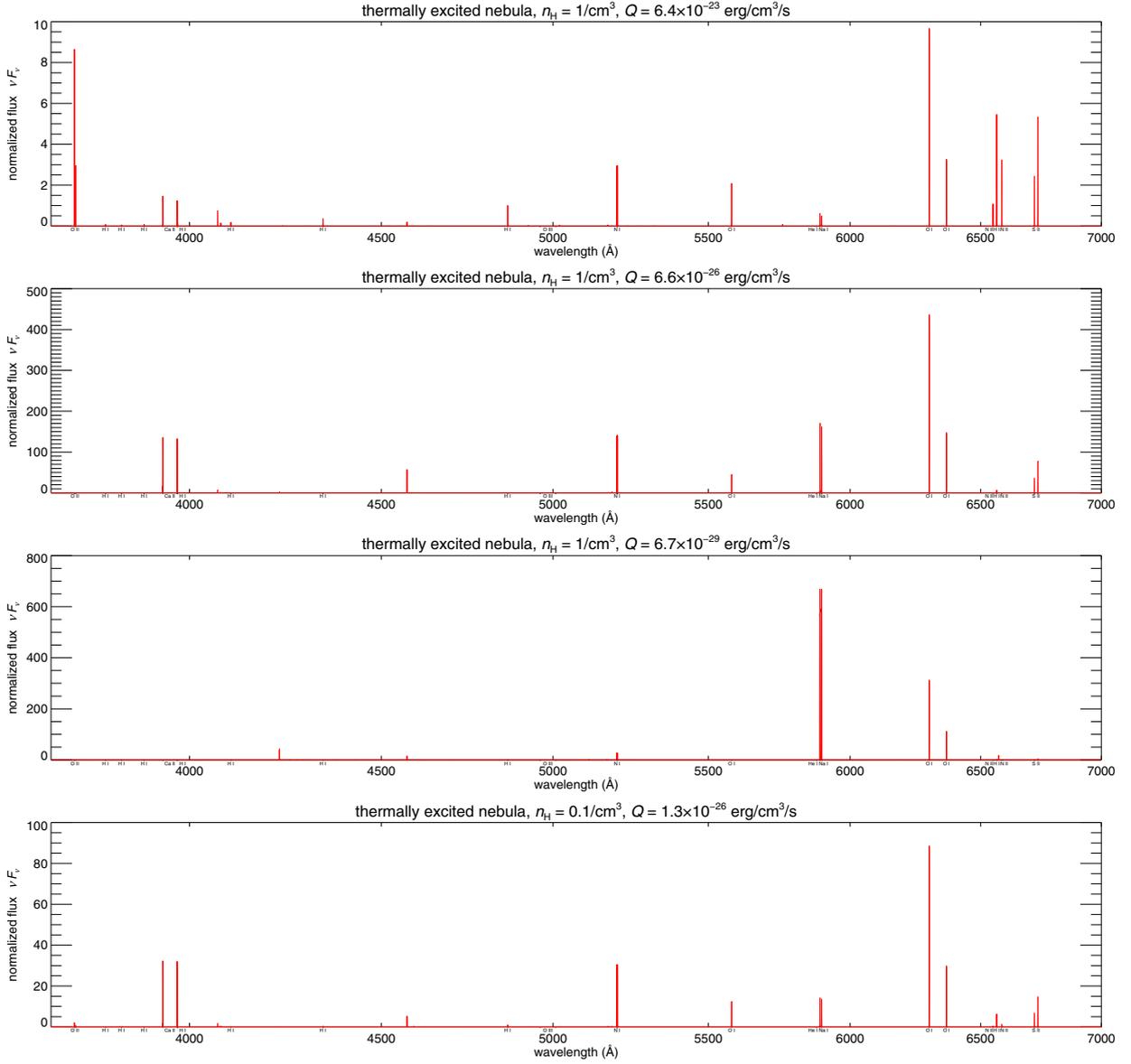

**Fig. 10.** Spectra of models without photoionization, excited by a uniformly distributed heat source. (Models B4a, B4b, B4c, C4c (from top to bottom); the parameters of the models are listed in Table 3).

The temperature, electron density, and ionization fractions turn out to be constant over the entire bubble, with a sharp cut-off at the outer border corresponding to the edge of the heated area. This means that the physical state of the gas is determined almost exclusively by *local* processes (heating and cooling), and that the intensity of ionizing radiation is either insignificant, or the ionizing radiation, if it exists, is not transported from grid point to grid point. The optical spectra thus represent those local physical properties, with negligible influence of the boundary region.

Indeed, in these models, the intensity of ionizing radiation is negligible (see Fig. 13 compared to Fig. 12; we plot the intensity times $r^2$ to compensate for the geometrical dilution of the radiation field; note also that these figures show the continuum only, as the lines can have intensities several orders of magnitude larger). Because the model shown still has a significant fraction of neutral hydrogen, any Lyman continuum radiation emitted by a recombination process is immediately reabsorbed. This is in contrast to the photoionized model where hydrogen is almost completely $H^+$; the inside of the nebula is therefore quite transparent to Lyman continuum radiation, and the intensity (times $r^2$) can even increase compared to that emitted by the central ionizing source.

The major difference between photoionized and thermally excited models is in the ionization balance. At the low densities of gaseous nebulae, the radiation field is mostly decoupled from the kinetic temperature of the gas, and may easily be dominated by an external source. In this case, the radiation temperature can thus reach much higher values (30...40 kK) than the kinetic temperature of the gas (10...20 kK). The fact that the ionization balance is dominated by radiative bound-free processes, whereas the intensities of the collisionally excited lines are coupled to the local gas temperature, leads to the characteristic H II region spectra. (Compare, for example, the strength of the [O III] emis-sion in the "high-excitation" model B4 with the greatly reduced [O III] emission of the "low-excitation" model B3; in contrast, the [N II] and [S II] emission is almost unaffected.) In the purely thermally excited models, the ionization balance is determined by the (much lower) local kinetic temperature, so that ionization stages III and higher are practically absent. This holds true even for model B4a, whose total energy input (via heating) exceeds the total energy input (via ionizing radiation) into a photoionized region of the same size (model B4) by more than a factor of five (cf. Table 3).

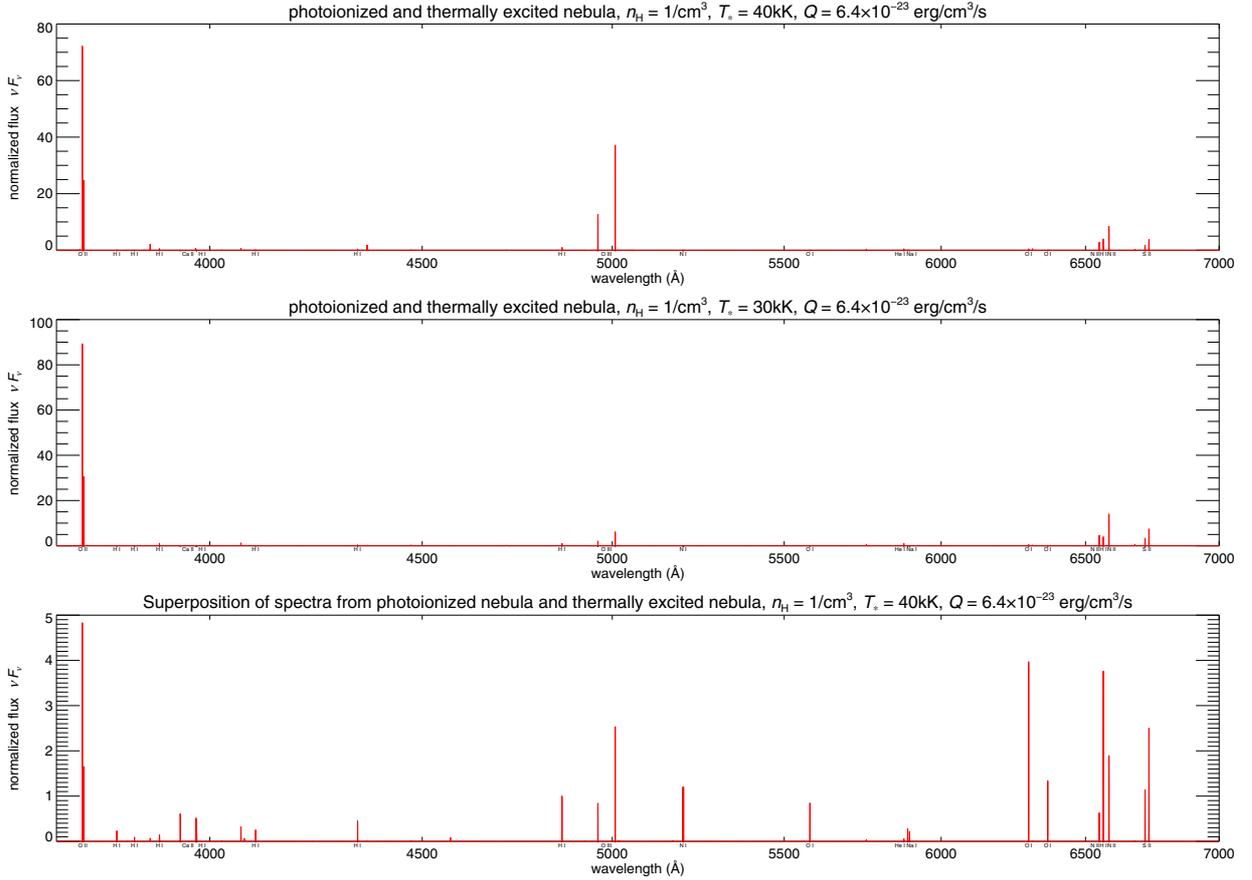

**Fig. 11.** Upper two panels: spectra of models heated through photoionization as well as a uniformly distributed heat source (models B4a1 and B3a1). Bottom panel: superposition of spectra from models B4 (photoionization only) and B4a (no photoionization, uniformly distributed heat source only).

A notable point regards the nebular temperature that ensues in equilibrium. In the photoionized nebulae, the temperature remains fairly constant over a wide range of nebular densities and effective temperatures of the ionizing source. This is due to the self-regulating mechanism of the main heating process, namely photoionization heating by hydrogen, which is proportional to the neutral hydrogen density, $n_{HI}$. Any increase in the heating via this process would also imply an increase in the ionization rate, leading to a reduced $n_{HI}$, thus counteracting the increased heating rate. No such regulation exists in thermally ionized plasma, and these exhibit a much larger range in temperatures depending on the given heating rate.

Models C4a and C4b fail to converge to a temperature within the limits of the radiative transfer code (several $10^4$ K) because the chosen heating rates exceed the cooling capacity of the gas at the given densities (cf. Dalgarno & McCray 1972) – at temperatures $T_e \gtrsim 15000$ K collisional excitation of neutral hydrogen becomes a major cooling process, but increasing the temperature further leads to a decrease in the density of neutral hydrogen via collisional ionization and therefore to a reduced cooling rate, and a thermal runaway occurs. (In reality, a new equilibrium would be reached at several $10^6$ K; our code, however, is not equipped to model gas at these temperatures.) This runaway does not happen in gas heated by photoionization, since there the heating rate is also proportional to the neutral hydrogen density.

Pure collisional ionization leads to significant ionization fractions only at comparatively high temperatures. For example, a hydrogen ionization fraction of 0.5 is only achieved at a temperature of about 15000 K (e.g., Sutherland & Dopita 1993, in agreement with our own calculations), whereas observed collisionally excited line ratios indicate much lower temperatures. Reynolds, Haffner, & Tufte (1999) estimate the temperature from the [N II]/H$\alpha$ ratio to be 7000 K to 12000 K. However, they base this estimate on the assumption that $N^+/N = H^+/H$ which is true for a collisionally ionized gas but not in general for a photoionized one, where $N^+$ is typically ionized to $N^{++}$ in the inner regions. The very low observed [O I] (and [N I], cf. Tüllmann & Dettmar 2000) intensities rule out that a significant fraction of the emitting gas is in these low ionization stages.

### 4.3. Combined photoionized and thermally excited models

Finally, in Fig. 11 we show the synthetic spectra from models where the photoionization has been augmented by heating from magnetic reconnection. We have used a constant additional heating rate (cf. Table 3) throughout the hydrogen Strömgren sphere, but we restrict the heating to that region. (Heating in the region outside the photoionized Strömgren sphere leads essentially to a superposition of emission lines from a purely thermally excited gas, with the corresponding prominent emission lines (see Fig. 11, bottom panel) of the lower ionization stages O I and N I, which are not observed. The additional heating rate must therefore be restricted to the photoionized region, although the detailed mechanism by which this is accomplished in nature is unclear at the moment. However, if the DIG H II is matter-bounded as opposed to radiation-bounded, then no extended neutral regions exist and the issue of such a restriction does not arise.) The additional energy input enhances the ionization rates and thus also increases the ionized region compared to the purely photoionized models; the radius within which we apply the additional heating has been adapted iteratively to correspond to the resulting Strömgren sphere. This increased volume allows a much larger amount of thermally injected energy to be processed in the region where the higher ionization stages are present. In particular we note that the energy input via heating into the ionized bubble of model B4a1 exceeds that of radiation by a factor of twenty.

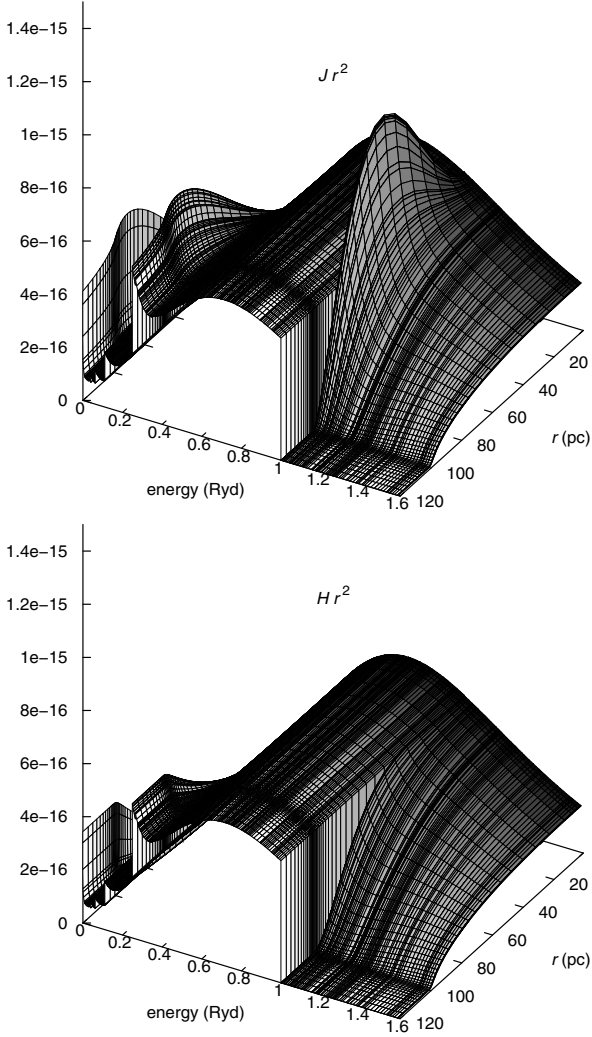

**Fig. 12.** Mean intensity (top) and flux (bottom) of the continuum radiation as function of radius and wavelength for an H II region ionized by UV radiation (model B4). The gas is transparent below 1 Ryd and can therefore ionize elements with ionization thresholds lower than that of hydrogen, such as Carbon and Sulfur, even outside the hydrogen Strömgren sphere. However, as the main heating source – photoionization of hydrogen – is not present in this outer region, the temperature is generally too low for significant collisionally excited line radiation to be generated there.

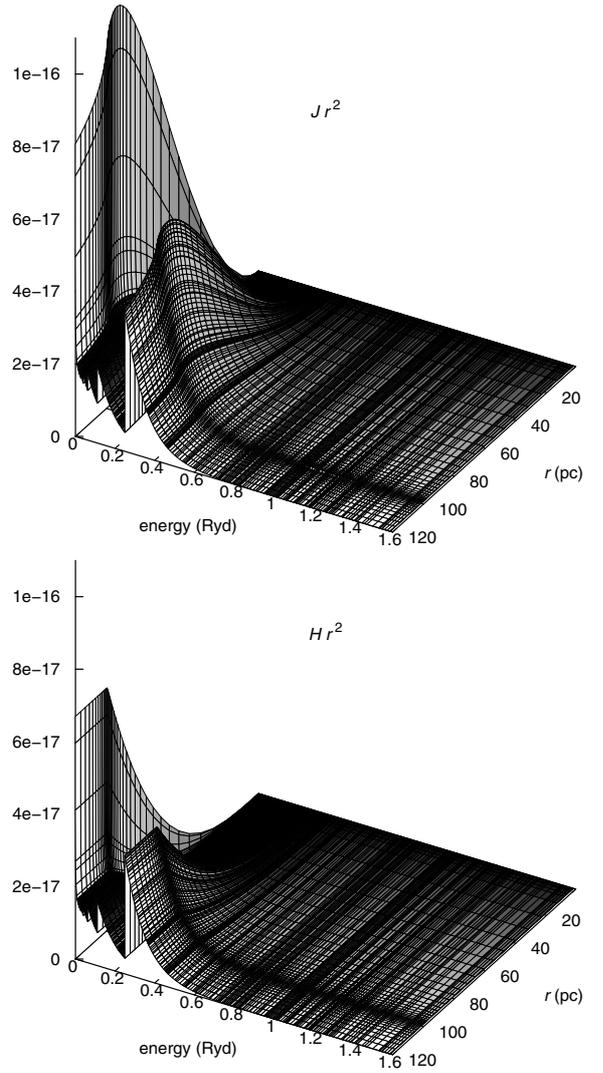

**Fig. 13.** Mean intensity (top) and flux (bottom) of the continuum radiation as function of radius and wavelength for a thermally excited nebula (model B4a).

The run of the ionization fractions as a function of the radius is similar to that of the photoionized models (i.e., a large fraction of oxygen and nitrogen is in the ionization stages O III and N III in the inner regions, with an increase of O II and N II toward the edge of the Strömgren sphere). The ionic species from which we observe emission lines are thus the same as in the purely photoionized models; however, being thermally excited, the strength of these lines is greater than from the purely photoionized models due to the higher temperature as a result of the additional heating rate. Given that we have used a fairly high heating rate in the models shown, intermediate line strengths between those from Fig. 11 (heated photoionized models) and Fig. 9 (purely photoionized models) may be achieved with a smaller heating rate.

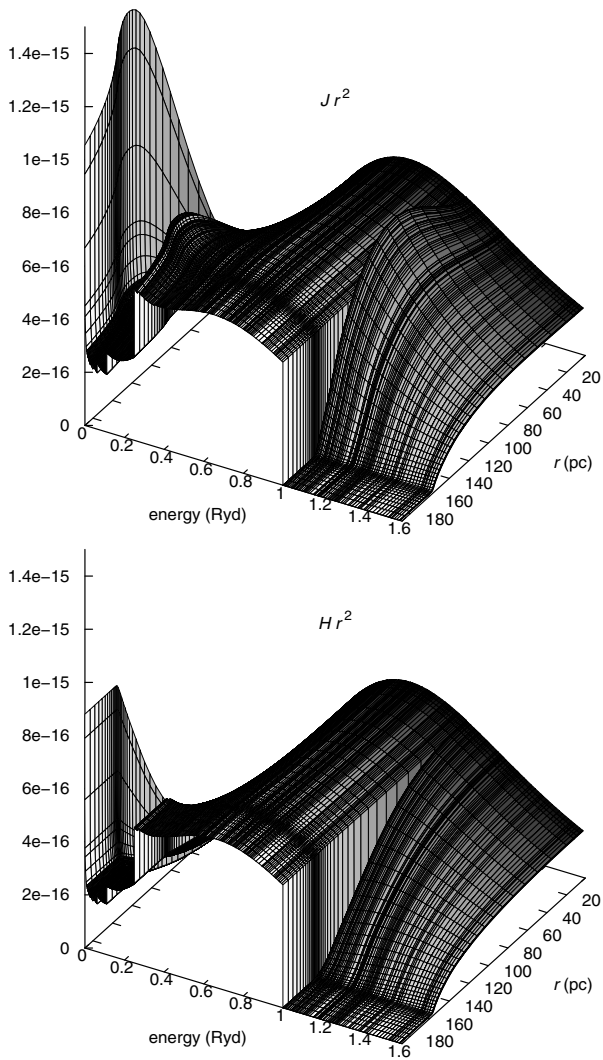

**Fig. 14.** Mean intensity (top) and flux (bottom) of the continuum radiation as function of radius and wavelength for the combined photoionized and thermally excited nebula (model B4a1).

## 5. Summary and conclusions

In this paper we have investigated the possibility of heating extraplanar ionized gas by ohmic dissipation in three-dimensional turbulent reconnective plasma–neutral-gas systems. By means of multi-fluid simulations including a current- or collision-driven resistivity model we have shown that a small fraction of the existing magnetic field strength suffices to provide a decisive supplemental heating rate. Our results indicate that the reconnection process on its own can generate enough power to ionize the DIG high above the galactic plane where photoionization by O and B stars has been argued to be insufficient, and may, depending on magnetic field strength and density, even contribute locally to X-ray emission, which has been observed in many halos of spiral galaxies where diffuse ionized gas is found (Dettmar 1998). Our radiative transfer simulations of the hot gas, however, show that the spectrum from a purely thermally heated medium is incompatible with the observed DIG spectra, and a photoionization source is required in addition to provide the higher ionization stages whose lines are observed. Combined heated and photoionized models, on the other hand, can account for the large [N II]/H$\alpha$, [S II]/H$\alpha$, and [O II]/[O III] observed line ratios.

*Acknowledgements.* We thank R.-J. Dettmar for providing the data of the observed spectra. This work was supported by the Deutsche Forschungsgemeinschaft under grant Pa 477/7-1.